\documentclass[aps,pra,showkeys,twocolumn,superscriptaddress]{revtex4-2}
\usepackage{amsmath, amssymb}
\usepackage{mathrsfs}
\usepackage{array}
\usepackage{booktabs}
\usepackage{tabu}
\usepackage{dcolumn}
\usepackage{amsmath}
\usepackage{amsfonts}
\usepackage{float}
\usepackage{amssymb}
\usepackage{graphicx,color}
\usepackage[colorlinks={true}]{hyperref}
\hypersetup{colorlinks=true,linkcolor=red,citecolor=blue,urlcolor=blue}
\usepackage{graphicx}
\usepackage{subfigure}
 \usepackage{multirow}
\usepackage{graphicx}
\usepackage{dcolumn}
\usepackage{bm}
\usepackage{pstricks}
\usepackage{braket}
\usepackage{orcidlink}
\begin{document}

\title{Spectral filtering and crystal length as control parameters for conditional correlations in quantum imaging}

\author{Hashir~Kuniyil\orcidlink{0000-0003-0338-1278}\thanks{Email: \texttt{hkuniyil@hbku.edu.qa}} Asad Ali\orcidlink{0000-0001-9243-417X} Mohammadreza Rezaee\orcidlink{0009-0007-4623-2113}\thanks{Email: \texttt{mrezaee@hbku.edu.qa}}   Saif Al-Kuwari\orcidlink{0000-0002-4402-7710}\thanks{Email: \texttt{smalkuwari@hbku.edu.qa}}}

\affiliation{Qatar Center for Quantum Computing, College of Science and Engineering, Hamad Bin Khalifa University, Doha, Qatar}
\date{\today}

\begin{abstract}
We establish spectral filtering as a control parameter for the conditional momentum and position correlations of SPDC biphotons used in quantum imaging. The conditional momentum uncertainty in spontaneous parametric down-conversion (SPDC) is strongly crystal-length and spectral-filter dependent along the walk-off axis; therefore, the effect is observed only in critically phase-matched (CPM) crystals such as $\beta$-barium borate (BBO) crystals, while quasi-phase-matched (QPM) crystals and the non-walk-off axis of BBO remain scaled strictly according to the standard pump waist size ($w_0$) dependent scaling law $1/w_0$- independent of the filter. In position space, the spectral-filter effect is universal and produces a flat–dip–rise (FDR) profile in every crystal class examined. Although this FDR profile was previously demonstrated in BBO only in the nondegenerate regime, our results establish its generality: the FDR dip is also present at exact degeneracy in QPM crystals, an unexpected feature that was previously thought to deliver a resolution advantage in the nondegenerate regime alone. Our treatment applies to all SPDC-based quantum-imaging regimes (including CPM and QPM crystals, walk-off and non-walk-off axes, degenerate and nondegenerate emission, and signal and idler filtering) and offers enhanced quantum-imaging resolution via the design rules presented—most prominently along the walk-off axis of CPM crystals (in far field) and across all transverse axes in the near field.
\end{abstract}

\maketitle
\section{Introduction}
Quantum imaging with biphotons has become one of the most versatile sources of quantum optics, with the transverse Einstein–Podolsky–Rosen (EPR) correlations produced by spontaneous parametric down-conversion (SPDC) standing out as a powerful resource \cite{howell2004realization, walborn2010spatial, kuniyil2021efficient, pearce2026quantum, law2004analysis}. This source supports a range of imaging modalities that surpass or circumvent classical constraints: resolution enhancement through biphoton centroid estimation \cite{toninelli2019resolution}, enhanced resolution using strong interference correlation, improved signal-to-noise ratio \cite{gregory2021noise,kuniyil2022noise}, sub-shot-noise imaging \cite{jedrkiewicz2004detection, brida2010experimental}, ghost imaging \cite{pittman1996two}, imaging with undetected photons \cite{lemos2014quantum, lahiri2015theory}, and elimination of the classical guide-star requirement for aberration correction \cite{cameron2024adaptive, zheng2025position}. More recently, the strong spatial correlation of SPDC has been used to recover lost images in turbulent media \cite{verniere2026entanglement}. In all of these schemes, the essential underlying resource is the transverse spatial correlation between the signal and idler photons, which enables a measurement on one photon to predict the corresponding conjugate property of its partner \cite{aspden2013epr, moreau2018ghost, joobeur1996coherence}. The resource fidelity is set by the conditional correlation uncertainty: the residual spread in one photon's transverse position or momentum once its partner is fixed. This conditional uncertainty is understood as a quantity that ultimately limits the achievable spatial resolution \cite{moreau2018resolution, fuenzalida2022resolution}, therefore controlling it is important to the design of any biphoton-based imaging system.

Conditional spatial uncertainty is governed primarily by two input parameters: the spatial waist $w_0$ of the pump beam \cite{pires2011type,fuenzalida2022resolution} and the length $L$ of the nonlinear crystal \cite{Schneeloch2016TransverseSPDC}, which is needed for SPDC generation. In far field (momentum space) imaging, the conditional momentum uncertainty ($\sigma_{q_i\mid q_s = 0}$) of a Gaussian-pumped source obeys the well known pump-limited scaling $\sigma_{q_i|q_s = 0}\approx 1/w_0$, independent of crystal length \cite{fuenzalida2022resolution}; in the near field (position space), the conditional position uncertainty instead broadens with crystal length through the phase-matching sinc function. We recently showed that in the nondegenerate regime, an appropriately chosen spectral-filter bandwidth tightens the conditional position correlation in BBO, improving the near-field spatial resolution by approximately $10\%$ \cite{Kuniyil2026entanglement}. In the study, however, the far-field conditional momentum uncertainty exhibited negligible dependence on the filter along the non-walk-off axis. Spectral filtering, by itself, is therefore not a useful control parameter for far-field correlation. This work shows that, similar to near field, the spectral filter dependent effect is attainable in momentum space but is observed only in walk off axis, and therefore observed only in critically phase-matched (CPM) nonlinear crystals such as BBO. We show that conditional uncertainties are jointly controlled with spectral filter bandwidth and nonlinear crystal length--- proving the advantage is confined to CPM crystals, in which the Poynting walk off is an operative parameter.  

Our investigation shows that the walk-off in CPM crystals contributes a term to the longitudinal phase mismatch parameter that is linear in the transverse momentum along the optic-axis (walk-off) plane. This term narrows the conditional momentum correlation below the pump-limited value $1/w_0$, and we analytically show that the narrowing increases with the crystal length as $\tfrac{1}{6}L^{2}\tan^{2}\!\rho$, where $\rho$ is the walk-off angle. The spectral configuration controls only as a modulator of this length-driven narrowing, through a single dimensionless ratio $R=\delta\lambda_{\mathrm{filter}}/\delta\lambda_{\mathrm{PM}}(\lambda,L)$ and a suppression function $\gamma(R)$: a narrow filter --- or operation near degeneracy, where $\delta\lambda_{\mathrm{PM}}$ is large --- preserves the narrowing ($\gamma\!\to\!1$), while a wide filter suppresses it out ($\gamma\!\to\!0$). The full dependence is captured in the closed form relation we introduced, whose suppression function $\gamma(R)$ collapses the data across PPKTP Type-0, PPKTP Type-I and BBO onto a common curve. At experimentally relevant lengths (up to $L=9$~mm for CPM crystals) the walk-off narrowing instead reaches $\sim\!9\%$ on the walk-off axis and becomes a usable resource. On every non-walk-off axis --- and hence for all quasi-phase-matched (QPM) crystals, where the walk-off term
is absent --- the conditional momentum uncertainty remains strictly pump-limited at $1/w_0$, independent of both $L$ and the filter.
 
In contrast to the momentum-space effect, which is selective to the walk-off axis of CPM crystals, the FDR profile observed in the position-space is universal. The conditional position uncertainty exhibits a non-monotonic dependence on the applied filter bandwidth—a flat plateau at narrow filtering, an intermediate minimum, and a rise at wide filtering (the flat–dip–rise, FDR, profile)—in every crystal class and on every transverse axis we examine. We first reported this profile in BBO in the nondegenerate regime~\cite{Kuniyil2026entanglement}. Here we show its generality, in particular, that the dip also appears under degenerate conditions. 
The presence of FDR profile in degeneracy is fundamentally different, and we show that the phase-matching width in the degenerate regime is governed by the group velocity dispersion (GVD), in contrast to the first-order group velocity mismatch (GVM) term that determines the SPDC bandwidth in the nondegenerate regime. We analyze this by Taylor-expanding the phase-matching function and directly show the dependence of GVM and GVD that makes the FDR profile differ in the nondegenerate and degenerate regimes. Furthermore, we establish that the FDR dip in the degenerate case is related by the equality relation $\delta\lambda_{PM} = \delta\lambda_{dip}$, where $\delta\lambda_{PM}$ SPDC's phase-matching bandwidth and $\delta\lambda_{dip}$ is the bandwidth at which the dip locatio is observed. This is in contrast to the nondegenerate case where a factor of 1.35 multiplies such as $\delta\lambda_{PM} = 1.35\delta\lambda_{dip}$, which we verified.  

The remainder of the paper is organized as follows. Section \ref{sec: model} presents the model used to investigate the conditional correlations in momentum and position space, with a spectral filter as a tunable parameter. Section \ref{sec:crystal_length_effect} develops an analytical framework to understand the effect of the walk-off dependent conditional momentum correlations variable as a function of crystal length and spectral filter width, and also provides details of the previously unreported scaling laws in momentum space. Section \ref{sec:fdr_ppktp} discusses the generality of the FDR profile observed in position-space imaging and, unique to this work, demonstrates the FDR profile in a degenerate regime of the QPM-based SPDC. Finally, section \ref{sec:dscussion} discusses the implications and applications of the finding and gives a summary of the findings.

\section{Model}\label{sec: model}
In an SPDC process, a high-frequency pump photon is down-converted into two lower-frequency daughter photons called the signal and idler. The spatial and spectral structure of the biphoton is governed by the phase-matching condition in the nonlinear crystal. In the paraxial approximation, the SPDC's phasematching can be understood using:
\begin{equation}
\Delta k_z \approx (k_p - k_s - k_i)
 - \frac{\lvert\mathbf q_s + \mathbf q_i\rvert^2}{2 k_p}
 + \frac{\lvert\mathbf q_s\rvert^2}{2 k_s}
 + \frac{\lvert\mathbf q_i\rvert^2}{2 k_i},
\label{eq:Deltak}
\end{equation}
where k is the wavevector, the indices p,s,i label the pump, signal and idler respectively, and $\mathbf q_{s,i}$ are the transverse momenta of the daughter photons.

For Type-I SPDC with BBO, the pump propagates as an extraordinary wave, making its Poynting vector tilted relative to its wavevector; this Poynting walk-off introduces an additional term proportional to the transverse momentum in the walk-off plane (the yz plane, taking the optical axis along z):
\begin{equation}
\Delta k_z^{\text{Type-I, BBO}} = \Delta k_z - (q_{s,y} + q_{i,y})\tan\rho,
\label{eq:Dk_BBO}
\end{equation}
where $\rho$ is the walk-off angle. In QPM crystals, this walk-off term is absent (the daughter photons propagate as principal-axis modes); instead, periodic poling of the nonlinear susceptibility contributes a discrete reciprocal-lattice vector $G_m = 2\pi m/\Lambda$ (m is an odd integer- indicates the order of the QPM crystals) to the longitudinal mismatch,
\begin{equation}
\Delta k_z^{\text{QPM}} = \Delta k_z - G_m,
\label{eq:Dk_QPM}
\end{equation}
with $\Lambda$ is the poling period. This work focuses on first-order QPM (m=1) throughout. The three crystals and Type considered in this work differ only by the polarization combinations of the three photons, which in turn fixes which principal index enters each wavenumber:
\begin{align}
k_{s,i}^{\text{Type-I, BBO}}  &= \frac{\omega_{s,i}}{c} \, n_o(\omega_{s,i})  && (e\to o+o), \label{eq:k_BBO}\\
k_{s,i}^{\text{Type-0, PPKTP}} &= \frac{\omega_{s,i}}{c} \, n_z(\omega_{s,i}) && (z\to z+z), \label{eq:k_T0}\\
k_{s,i}^{\text{Type-I, PPKTP}} &= \frac{\omega_{s,i}}{c} \, n_y(\omega_{s,i}) && (z\to y+y). \label{eq:k_T1}
\end{align}
The KTP principal indices $n_y$ and $n_z$ are taken from the Kato--Takaoka Sellmeier equations with the Emanueli--Arie thermo-optic correction~\cite{kato2002sellmeier, emanueli2003temperature}; the BBO indices $n_o,\,n_e$ are standard Sellmeier equations. The pump in all three cases uses $k_p = (\omega_p/c)\, n_z(\omega_p)$ (PPKTP) or $n_e(\omega_p, \theta_{\rm pm})$ (BBO). Within the paraxial approximation the transverse biphoton amplitude is the product of a Gaussian pump envelope and a sinc phase-matching factor:
\begin{equation}
\Phi(\mathbf q_s, \mathbf q_i; \lambda_s)
= \exp\left[-\frac{w_0^{2}}{4}(\mathbf q_s + \mathbf q_i)^{2}\right] 
\mathrm{sinc}\left(\frac{\Delta k_z L}{2}\right),
\label{eq:Phi}
\end{equation}
where $w_0$ is the size of the pump waist, L is the length of the crystal, and the appropriate $\Delta k_z$ from Eqs.~\eqref{eq:Dk_BBO}--\eqref{eq:Dk_QPM} is used for each class of crystals.
\paragraph*{Estimating conditional uncertainty.}
Estimation of conditional uncertainties requires a point-to-point correlation analysis of the spatial field. We used the position-conditioned point estimator method to evaluate the conditional correlation in both near-field and far-field imaging. This method is as follows: fix one photon at a chosen reference point in the transverse plane and read the distribution of its partner. Mathematically, this estimator is written as follows.
\begin{equation}
  \Delta q_{s|i} \;\equiv\; \sigma_{q_s|q_i=0}
  = \left[\langle q_s^{2}\rangle_{q_i=0} - \langle q_s\rangle_{q_i=0}^{2}\right]^{1/2},
  \label{eq:Delta-q}
\end{equation}
and in the near field with $\mathbf q \to \mathbf x$. Operationally, $\sigma_{q_s \mid q_i=0}$ is extracted by fitting the row of $|\Phi(\mathbf q_s,\mathbf q_i)|^{2}$ at $\mathbf q_i = 0$ to a Gaussian $A\exp[-(q_s-\mu)^{2}/(2\sigma^{2})]$, and likewise in position space.

We apply a Gaussian-shaped band-pass filter with full width at half maximum (FWHM) equal to \(\mathrm{FWHM}\), centered at \(\lambda_s^{(0)}\). The filter transmission as a function of the signal wavelength \(\lambda_s\) is
\begin{align}
 F_s(\lambda_s) = \exp\left[-\frac{(\lambda_s - \lambda_{s0})^2}{2\sigma_\lambda^2}\right],   
\end{align}

where \(\lambda_{s0}\) is the center wavelength of the filter and \(\sigma_\lambda\) is the standard deviation of the Gaussian, related to the FWHM by
\begin{align}
    \sigma_\lambda = \frac{\mathrm{FWHM}}{2\sqrt{2\ln 2}}.
\end{align}
With this filter applied, the spectrally averaged biphoton amplitude is given by 
\begin{equation}
  I(\mathbf q_s, \mathbf q_i; \lambda_s)
  = \int d\lambda_s\, |F(\lambda_s)|^{2}\, |\Phi(\mathbf q_s,\mathbf q_i;\lambda_s)|^{2}.
  \label{eq:Iq}
\end{equation}
This corresponds to the far-field biphoton distribution when a spectral filter is applied downstream of the signal photons. For applying the filter in the idler arm, an adjustment of the parameter values would be needed. 

The near-field joint distribution is the Fourier conjugate of the momentum-space amplitude and, in any realistic detection scheme, is also averaged over the pass-band of a spectral filter placed in the signal arm (idler arm if the filter is applied in the idler arm). 
\begin{align}
I(\mathbf x_s, \mathbf x_i; \lambda_s)
&= \int d\lambda_s\, |F(\lambda_s)|^{2}\nonumber\\ 
&\quad\times \left|\mathcal F^{-1}_{(\mathbf q_s,\mathbf q_i)\to(\mathbf x_s,\mathbf x_i)}
\bigl[\Phi(\mathbf q_s,\mathbf q_i;\lambda_s)\bigr]\right|^{2}.
\label{eq:Ix}
\end{align}
In near field spectral filtering based FDR analysis, the phasematching condition is modified into a convenient form as 
\begin{align}
    \Delta k(\Omega) = \Delta k_z^{(0)} - G_m-\tau\Omega-\frac{1}{2}k_{eff}^{''}\Omega^2,
    \label{eq:near_field_phasematching}
\end{align}
where $\Omega$ is the detuning frequency defined as $\Omega = \omega - \omega^{(0)}$. Detailed derivation for Eq.~\eqref{eq:near_field_phasematching} is given in the Appendix.\ref{app:dlPM_derivation}.
Numerically, Eqs.~\eqref{eq:Iq} and \eqref{eq:Ix} are implemented by sampling the grid $\mathbf q_{s,i}$$N\times N$ and performing, respectively, a 2D inverse FFT and a direct evaluation for each filter sample.
\section{Results A --- Crystal-length scaling of $\Delta q_{s|i}$
with spectral filtering}
\label{sec:crystal_length_effect}

\begin{figure*}
\centering
\includegraphics[width=0.95\linewidth]{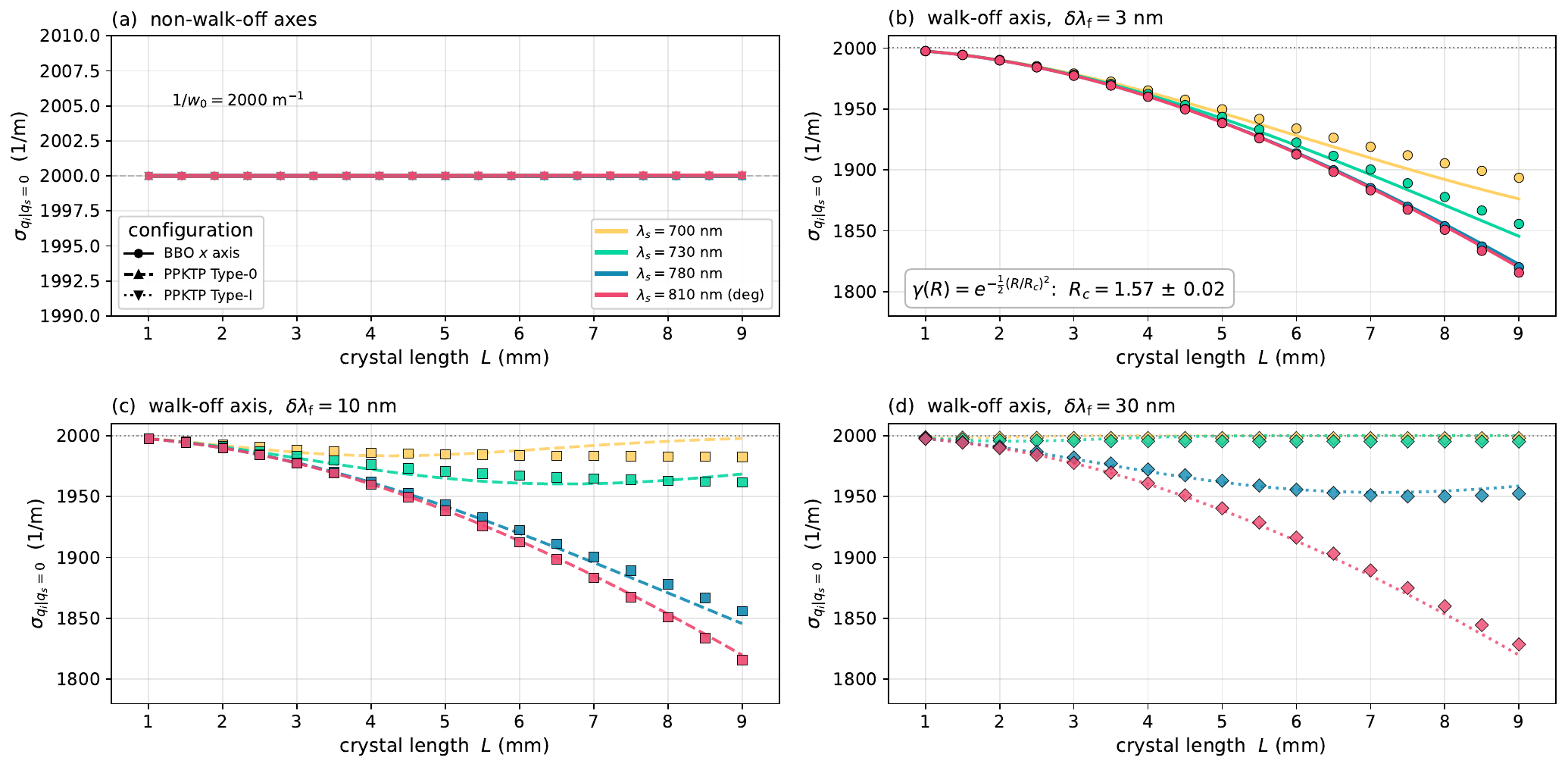}
\caption{Conditional momentum slice width $\sigma_{q\mid q_s=0}$ versus crystal length $L$ for four signal wavelengths at fixed $w_0 = 500\,\mu\mathrm{m}$ and signal-arm filter FWHM $\delta\lambda_{\rm filter} = \{3, 10, 30\}\,\mathrm{nm}$. 
\textbf{(a)} No-walk-off configurations (PPKTP Type-0, circles; Type-I, squares; BBO $x$ axis, triangles) collapse onto the pump-set plateau $1/w_0 = 2000\,\mathrm{m^{-1}}$, independent of both $L$ and $\delta\lambda_{\rm f}$. 
\textbf{(b–d)} BBO walk-off ($y$) axis. The walk-off term in $\Delta k_z$ produces an $L$-dependence controlled by the filter through the dimensionless ratio $R$ [Eq.~\eqref{eq:R}] at different filter bandwidths: (b) $\delta\lambda_{\rm f} = 3\,\mathrm{nm}$, (c) $\delta\lambda_{\rm f} = 10\,\mathrm{nm}$, and (d) $\delta\lambda_{\rm f} = 30\,\mathrm{nm}$.}
\label{fig:dq_vs_L}
\end{figure*}

The analysis of the SPDC biphoton amplitude presented in section \ref{sec: model} is carried out to understand the conditional momentum correlation (Eq.~\eqref{eq:Iq}) for a fixed pump beam waist set at $w_0 = 500\,\mu\mathrm{m}$, while varying the filter bandwidth and crystal length. As shown in Fig.~\ref{fig:dq_vs_L} (a), a comparison of the four SPDC configurations shows that $\sigma_{q|q_s = 0}$ is invariant with respect to crystal length in the absent walk off axes, which include the type 0 and type 1 PPKTP and the non-walk-off axis of the BBO crystal. As seen, all non–walk-off axes strictly obey the standard and well-known pump beam waist dependence, namely $\sigma_{q|q_s = 0} \propto 1/w_0$. Although we have not explicitly tested on type II, this effect is expected to be held as this is an axis based effect. The walk-off axis of the critically phasematched crystal exhibits a change dependent on the crystal-length and filter-bandwidth in $\sigma_{q|q_s = 0}$. Figure~\ref{fig:dq_vs_L}(b) -(d) shows the degenerate and nondegenerate $\sigma_{q|q_s = 0}$ as a function of the crystal length. The three filter bandwidths tested are $\delta\lambda_f \in \{3\,\mathrm{nm}, 10\,\mathrm{nm}, 30\,\mathrm{nm}\}$, were applied to four SPDC daughter-photon wavelength combinations: $780\,\mathrm{nm}$, $730\,\mathrm{nm}$, and $700\,\mathrm{nm}$, together with the degenerate case at $810\,\mathrm{nm}$ (all configurations assume a single-frequency pump with a central wavelength of $\lambda_p = 405\,\text{nm}$.). The degenerate case shows a strong crystal-length dependence, and when a narrow filter is applied, the nondegenerate cases tend to approach the same trend as the degenerate case. We investigate this effect by following the analytical treatment in the next section.

\subsection{conditional momentum correlation uncertainty at $q_s = 0$}
\label{ssec:slice_amp}

Setting $q_s=0$ in the paraxial biphoton amplitude
[Eq.~\eqref{eq:Phi}] gives
\begin{equation}
|\Phi(0,q_i;\lambda_s)|^{2} = \exp\!\left[-\frac{w_0^{2}q_i^{2}}{2}\right]
\,\mathrm{sinc}^{2}\!\left[\frac{L\,\Delta k_z(0,q_i;\lambda_s)}{2}\right],
\label{eq:slice_amp}
\end{equation}
with the longitudinal mismatch expanded to the second order in $q_i$,
\begin{equation}
\Delta k_z(0,q_i;\lambda_s) = \Delta k_0(\lambda_s) + \alpha\,q_i + \beta\,q_i^{2}.
\label{eq:Dkz_q}
\end{equation}
$\Delta k_0$ is the collinear mismatch (zero at the design wavelength), $\beta=(k_p-k_i)/(2 k_p k_i)$ is the Fresnel coefficient common to all geometries, and
\begin{equation}
\alpha =
\begin{cases}
0, & \text{PPKTP Type-0, Type-I, BBO $x$ axis},\\[2pt]
-\tan\rho, & \text{BBO $y$ axis (walk-off direction).}
\end{cases}
\label{eq:alpha_cases}
\end{equation}
Simply, $\alpha = 0$ whenever there is no walk off effect. Therefore, this is also applicable to non-walk axis of type -II SPDC, which is not explicitly tested in this work. If the walk off axis is considered, regardless of the crystal, the walk of angle $\rho$ of the particular crystal should be considered. 

\subsection{No-walk-off case: dependence of conditional momentum uncertainty exactly at $1/w_0$}
\label{ssec:nowalkoff}

The sinc argument is $|L\beta q_i^{2}/2|$ with first zero $q_i^{\rm sinc}\!\approx\!10^{5}\,\mathrm{m^{-1}}$, two orders of magnitude larger than $1/w_0$, where $w_0 = 500\,\mu m$.  The sinc factor is unity with $<10^{-6}$ across the fixed pump; leading Eq.~\eqref{eq:slice_amp} reduces to a pure Gaussian, and yielding the well known scaling law in the far-field measurement \cite{Schneeloch2016TransverseSPDC}
\begin{equation}
\sigma_{q_i|q_s=0}^{\,(\alpha=0)} = \frac{1}{w_0}.
\label{eq:sigma_nowalkoff}
\end{equation}
This concludes conditional momentum correlation is a variant to only pump beam waist, a strictly stable parameter that unchanged with crystal length, nondegenerate and degenerate conditions and filter bandwidth.
\subsection{Walk-off case, degenerate limit: closed form}
\label{ssec:walkoff_degen}
On the BBO walk-off axis, the linear term dominates the sinc at the pump scale (quadratic term ineffective due to previously explained reasons). At the design wavelength $\Delta k_0 = 0$, and Taylor-expanding $\mathrm{sinc}^{2}(L\alpha q_i/2)$ to second order in $q_i$ followed by matching to a Gaussian (ignoring the second order term in Eq.\eqref{eq:Dkz_q}),
\begin{equation}
\mathrm{sinc}^{2}\!\left(\frac{L\,\alpha\,q_i}{2}\right)
\approx 1 - \frac{(L\alpha q_i)^{2}}{12}
\;\approx\; \exp\!\left[-\frac{1}{2}\frac{L^{2}\tan^{2}\!\rho}{6}\, q_i^{2}\right],
\label{eq:sinc-approx}
\end{equation}
combining with the pump factor yields a single Gaussian whose inverse-square width is the sum of two contributions (one from the pump Gaussian and the other from a walk-off inherited term in the sinc function). The degenerate closed-form slice width ibecomes
\begin{equation}
\sigma_{q_i|q_s=0}^{\,({\rm BBO}\,y),\,\varepsilon=0}(L)
\;=\;
\frac{1}{\sqrt{w_0^{2} + \tfrac{1}{6}\,L^{2}\tan^{2}\!\rho}}\,.
\label{eq:sigma_q_degenerate}
\end{equation}
For $L\tan\rho \ll w_0$ this asymptotically reaches $1/w_0$; for $L\tan\rho \gg w_0$ it scales to $\sqrt{6}/(L\tan\rho)$, falling to
$L^{-1}$. The crossover lies at the geometric scale $L_{\rm cross} = \sqrt{6}\,w_0/\tan\rho \approx 19.7$\,mm for our set parameters, hence the realistic experimental range (for BBO) $L = 1\text{--}9$\,mm lies at the prior place of the transition. After incorporating the parameters of $\lambda_s = 810$\,nm, $L = 9$\,mm into Eq.~\eqref{eq:sigma_q_degenerate} gives $\sigma \approx 1819\,\mathrm{m^{-1}}$, in agreement with the numerical $1816\,\mathrm{m^{-1}}$ to $0.2\%$, validating our analytical explanation with the simulated one and support with known relation in special conditions. The general trend of crystal length influenced change in conditional momentum correlation is shown in Fig.~\ref{fig:dq_vs_L} (b)-(d). In all figures, the analytical fit perfectly follows the simulated trend in the degenerate case. With $L = 9\,mm$, improve $\sigma_{q_i|q_s = 0}$ by approximately $9\%$, an advantage offered only the measurement on the walk-off axis of the BBO crystal. In the nondegenerate case, the unavoidable wavelength distribution undoes this behavior with magnitude of non-degeneracy as as discussed in the next section, Sec.\,\ref{ssec:walkoff_universal}.

\subsection{General expression for conditional momentum uncertainity}
\label{ssec:walkoff_universal}

In the nondegenerate regime, the spectral filter samples non-zero-mismatch sidebands; the incoherent sum undoes the walk-off-driven narrowing advantage (in biphoton based imaging). Therefore, we can expect a higher nondegeneracy and a higher reduction of $\sigma_{q_i\mid q_s}$ as seen in Figs.\,\ref{fig:dq_vs_L} (b) - (d). One can assume that a spectral filter can reduce this incoherent sum. A better way to test this effect is using narrow spectral filter with centered at wavelength of downconverted photon pairs. That can be studied introducing a dimensionless suppression
factor $\gamma(R)\in[0,1]$ with control parameter influenced by the spectral filter Eq.~\eqref{eq:sinc-approx}:
\begin{equation}
\sigma_{q_i|q_s=0}^{\,(\text{BBO}\,y)}(L,\lambda_s,\delta\lambda_{\rm filter})
=
\frac{1}{\sqrt{w_0^{2}+\tfrac{1}{6}\,\gamma(R)^{2}\,L^{2}\tan^{2}\!\rho}}.
\label{eq:sigma_walkoff}
\end{equation}
We observe that the control variable is the filter-to-PM bandwidth ratio (R), given by
\begin{equation}
R = \frac{\delta\lambda_{\rm f}}{\delta\lambda_{\rm PM}(\lambda, L)},
\qquad
\gamma(R) = \exp\!\left[-\tfrac{1}{2}(R/R_c)^{2}\right],
\label{eq:R}
\end{equation}
where $R_c$ is a fit parameter and we found the value of it $R_c\!\approx\!1.57\pm0.02$. The function of $\gamma(R)$ is Gaussian, as it is inherited from the functional form of the filter. Operationally, $R_c$ is the R value at which the walk-off induced suppression reaches the $1/\sqrt{e}$ level. The analytical expression we derived fit with simulation and observed it follows the trend as shown in Fig\ref{fig:dq_vs_L} (b)-(c). The figure shows the narrower filters limit the incoherent mixture of wavelengths--thereby improved conditional uncertainity. The phase-matching bandwidth $\delta\lambda_{\rm PM}(\lambda,L)$ is derived from the crystal Sellmeier and presented in Appendix~\ref{app:dlPM_derivation}; for the three crystal configurations of this work, reference values are tabulated in Table~\ref{tab:tab1}.  The wavelength $\lambda$ appearing in $\delta\lambda_{\rm PM}$ is the filtered photon's wavelength --- signal or idler depending on which arm carries the filter (Sec.~\ref{ssec:idler_arm}).

\subsection{Filter in the signal arm vs filter in the idler arm}
\label{ssec:idler_arm}

The dimensionless ratio $R$ depends on which arm the filter is placed. Using the closed form of the GVM derived in Appendix~\ref{app:dlPM_derivation}, we achieve 
\begin{equation}
\delta\lambda_{\rm PM}^{(s)}(\lambda_s,L) =
\frac{5.566\,\lambda_s^{\,2}}{2\pi\,L\,|n_g(\lambda_s)-n_g(\lambda_i)|}\,,
\label{eq:dlPM_signal}
\end{equation}
where $n_g(\lambda)$ is the group velocity refractive index for light with wavelength $\lambda$, the corresponding bandwidth seen by a filter on the idler arm can be obtained by replacing the prefactor $\lambda_s^{\,2}$
with $\lambda_i^{\,2}$, since the GVM denominator $|n_g(\lambda_s)-n_g(\lambda_i)|$ is symmetric under the signal--idler exchange.  With this, the two bandwidths are related by
\begin{equation}
\delta\lambda_{\rm PM}^{(i)}(\lambda_i,L)
= \left(\frac{\lambda_i}{\lambda_s}\right)^{\!2}\,
\delta\lambda_{\rm PM}^{(s)}(\lambda_s,L)
\label{eq:idler_scaling}
\end{equation}
For our four signal wavelengths at $\lambda_p=405$\,nm the geometric prefactor is $(\lambda_i/\lambda_s)^{2} = 1.885, 1.553,
1.378, 1.166$ for $\lambda_s = 700, 730, 750, 780$\,nm respectively. This means an idler-arm filter sees a phase-matching window 14--88\% wider than a signal-arm filter for the same physical pair--supporting the recently found scaling relation in near field imaging analysis \cite{Kuniyil2026entanglement} and the entanglement certification correction parameter \cite{brambilla2025}.

This is the same scaling relation that controls the FDR dip position in the near-field (Sec.~\ref{ssec:dip_scaling}); both
phenomena ultimately follow from the $\lambda^{2}$ prefactor in the frequency-to-wavelength conversion. Operationally, if the filter is placed in the idler arm, the dimensionless ratio becomes
\begin{equation}
R^{(i)} \;=\;
\frac{\delta\lambda_{\rm filter}}{\delta\lambda_{\rm PM}^{(i)}}
\;=\;
\left(\frac{\lambda_s}{\lambda_i}\right)^{\!2}
\frac{\delta\lambda_{\rm filter}}{\delta\lambda_{\rm PM}^{(s)}}
\;=\;
\left(\frac{\lambda_s}{\lambda_i}\right)^{\!2}\,R^{(s)}\,,
\label{eq:R_idler_arm}
\end{equation}
and the same $\gamma(R)$ of Eq.~\eqref{eq:R} is evaluated at $R^{(i)}$ instead of $R^{(s)}$.  Substituting back into Eq.~\eqref{eq:sigma_walkoff} gives the prediction for the idler-filtered curve directly.

The qualitative consequence for the crystal-length scan is that $\gamma(R^{(i)})$ saturates more slowly with $\delta\lambda_{\rm filter}$
than $\gamma(R^{(s)})$, because $R^{(i)} < R^{(s)}$ at the same filter width.  For a fixed filter, the suppression of the walk-off slope is therefore weaker when the filter sits on the idler arm, and the $\sigma_q(L)$ curves are accordingly steeper. This is consistent with the qualitative expectation that filtering the longer-wavelength photon delivers less spectral-averaging blurring than filtering the shorter-wavelength one.

A comparison is shown in Table~\ref{tab:arm_compare} for $L = 3$\,mm: $\delta\lambda_{\rm PM}^{(i)}$ is wider than $\delta\lambda_{\rm PM}^{(s)}$ by exactly the predicted $(\lambda_i/\lambda_s)^{2}$ factor for all three crystal classes and all four signal wavelengths.

\begin{table}[h]
\centering
\caption{Signal-arm vs.\ idler-arm $\delta\lambda_{\rm PM}$ at
$L=3$\,mm, $\lambda_p=405$\,nm.  The ratio is exactly
$(\lambda_i/\lambda_s)^{2}$ for every entry.}
\label{tab:arm_compare}
\small
\begin{tabular}{lcccc}
\hline
crystal & $\lambda_s/\lambda_i$ (nm) &
$\delta\lambda_{\rm PM}^{(s)}$ (nm) &
$\delta\lambda_{\rm PM}^{(i)}$ (nm) &
$(\lambda_i/\lambda_s)^{2}$ \\
\hline
BBO ($n_o$)      & 700/961  &  9.03 & 17.02 & 1.885 \\
                 & 730/910  & 14.07 & 21.84 & 1.553 \\
                 & 750/880  & 20.32 & 28.01 & 1.378 \\
                 & 780/842  & 45.70 & 53.30 & 1.166 \\
PPKTP T0 ($n_z$) & 700/961  &  2.41 &  4.55 & 1.885 \\
                 & 730/910  &  3.77 &  5.86 & 1.553 \\
                 & 750/880  &  5.46 &  7.53 & 1.378 \\
                 & 780/842  & 12.30 & 14.35 & 1.166 \\
PPKTP T1 ($n_y$) & 700/961  &  3.32 &  6.26 & 1.885 \\
                 & 730/910  &  5.19 &  8.06 & 1.553 \\
                 & 750/880  &  7.51 & 10.35 & 1.378 \\
                 & 780/842  & 16.91 & 19.73 & 1.166 \\
\hline
\label{tab:tab1}
\end{tabular}
\end{table}

With $\delta\lambda_{\rm PM}^{(\rm arm)}$ chosen according to which photon carries the filter and $\gamma(R)$ from Eq.~\eqref{eq:R}, reproduces the numerical slice data of Fig.~\ref{fig:dq_vs_L}.

\section{Results B --- The flat--dip--rise profile in PPKTP
across degenerate and nondegenerate regimes}
\label{sec:fdr_ppktp}

\begin{figure*}
\centering
\includegraphics[width=0.95\linewidth]{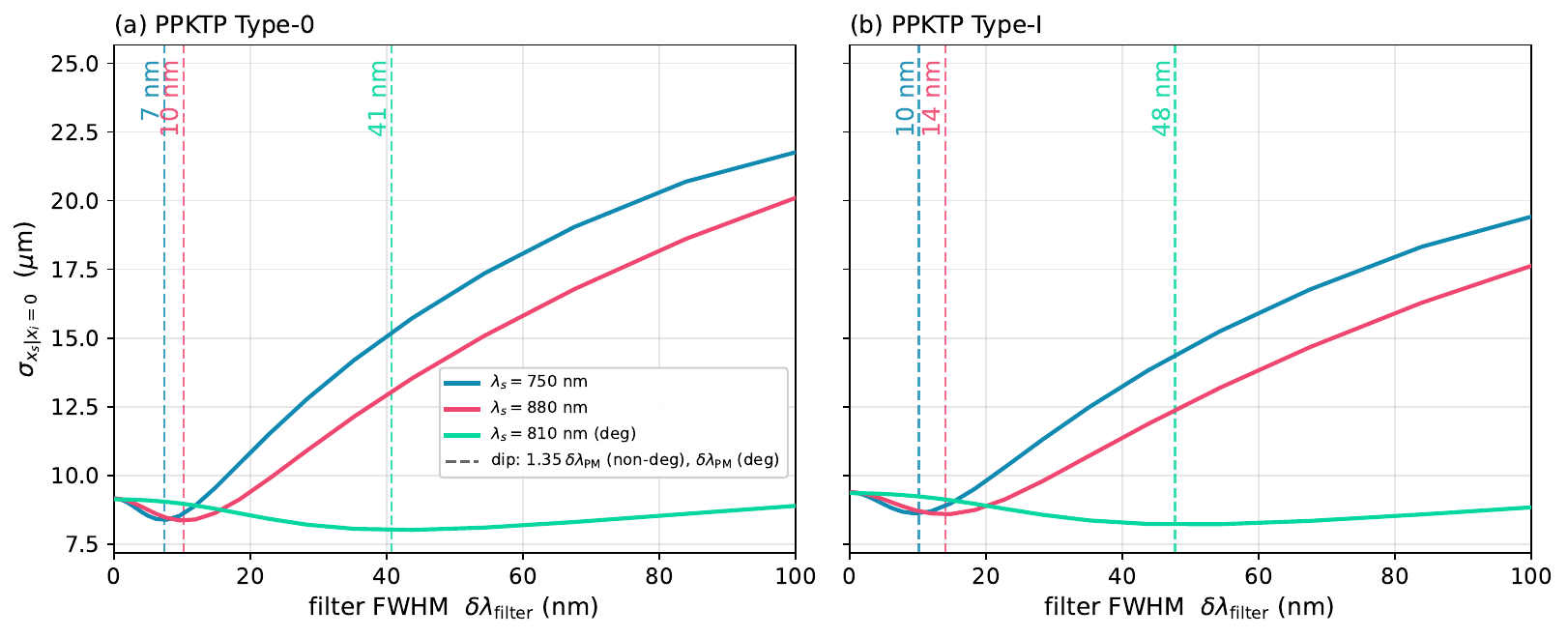}
\caption{Near-field conditional position uncertainty $\sigma_{x_s|x_i=0}$ as a function of the signal-arm spectral filter FWHM $\delta\lambda_{\rm filter}$ in (a) PPKTP Type-I and (b) PPKTP Type-0, at fixed crystal length $L = 3$\,mm, pump waist $w_0 = 500\,\mu$m, and pump wavelength $\lambda_p = 405$\,nm. Three signal wavelengths are plotted in each panel; both crystals show a FDR profile at every signal wavelength, including the degenerate 810 nm case.  Vertical markers in each panel illustrate the dip-position scaling Eq.~\eqref{eq:dip_scaling} for the $\lambda_s = 750$\,nm $\to \lambda_i = 880.4$\,nm conjugate pair.}
\label{fig:fdr_ppktp}
\end{figure*}

The FDR profile is the position-space conditional uncertainty $\sigma_{x_s|x_i=0}$ plotted against the spectral filter bandwidth $\delta\lambda_{\rm filter}$; it consists of a flat plateau insensitive to the spectral filter at the narrow filter (where $\sigma_{x_i|x_s=0} = \sigma_{\rm narrow}$ is set by the pump waist and crystal length alone), a dip at the intermediate filter where the spectral averaging tightens the joint along the anti-diagonal, and a rise at the wide filter where the incoherent averaging spreads the joint along the diagonal.  We previously identified this profile in BBO~\cite{Kuniyil2026entanglement} and characterized its distribution between the walk-off and non-walk-off transverse axes, and previously shown FDR profile measurement axis independent. This section reports two new observations in QPM crystals (PPKTP Type-0 and Type-I) that go beyond the  CPM crystals phenomenology.

\subsection{Filter dependence of the FDR profile in QPM crystals}
\label{ssec:fdr_observed}
Figure~\ref{fig:fdr_ppktp} shows the conditional position width $\sigma_{x_s|x_i=0}$ versus the filter FWHM in PPKTP Type-I (panel a of the figure) and Type-0 (panel b) for six signal wavelengths spanning the strongly nondegenerate range ($\lambda_s = 700$ \,nm, $\lambda_i = 961$\,nm) to degenerate range ($\lambda_s = \lambda_i = 810$ \,nm) at fixed $L = 3$\,mm and $w_0 = 500\,\mu$m.  Unlike the far-field momentum correlation of Sect.~\ref{sec:crystal_length_effect}, which shows filter dependence only in the BBO walk-off geometry, the near-field FDR profile is general, and present in all crystal configuration: both PPKTP Type-0 and Type-I.  The flat-plateau baseline is $\sigma_{\rm narrow}\approx 9.3\,\mu$m for Type-I and $\approx 9.1\,\mu$m for Type-0, set by the pump waist and crystal length. The dip depth is ${\sim}10\%$ of the plateau in the nondegenerate cases (recovering the BBO observation).  The novel feature is that the degenerate curve ($\lambda_s = 810$ \,nm) in both PPKTP types displays a clear dip at filter widths near $\delta\lambda_{\rm filter} \approx 50$\,nm --- inside the experimentally accessible range and deeper than the nondegenerate dips (hence QPM degenerate can provide resolution advantage; a feature previously thought only present in the nondegenerate regimes).

\subsection{Dip-position scaling relation}
\label{ssec:dip_scaling}

For nondegenerate signal--idler pairs, the dip filter width on the signal arm and the dip filter width on the idler arm are related by
the same geometric scaling we derived in Sec.~\ref{ssec:idler_arm}:
\begin{equation}
\delta\lambda^{\,\rm dip}_{\rm idler}
= \left(\frac{\lambda_i}{\lambda_s}\right)^{\!2}
\delta\lambda^{\,\rm dip}_{\rm signal}\,.
\label{eq:dip_scaling}
\end{equation}
The relation follows from the dip filter width being $R_c$ times $\delta\lambda_{\rm PM}$, which carries the $\lambda^2$ prefactor of the filtered photon while sharing the GVM denominator with its conjugate. As a result, Eq.~\eqref{eq:dip_scaling} is an analytical relation that does not depend on the underlying crystal Sellmeier in any way: any second-order corrections to the absolute $\delta\lambda^{\rm dip}$ affect both arms by the same multiplicative factor and cancel in the ratio.

For the conjugate pair $(\lambda_s, \lambda_i) = (750, 880.4)$\,nm the predicted scaling factor is $(880.4/750)^{2} = 1.378$.  The
markers in Fig.~\ref{fig:fdr_ppktp} compare this prediction with the numerical dip filter width of the 880 nm column in each panel. Table~\ref{tab:fdr_reduction} lists the corresponding filter widths in dip and fractional reductions in the conditional position width for all six wavelengths in both types of crystal.

\subsection{The degenerate dip in QPM crystals}
\label{ssec:degenerate_dip_ppktp}
\begin{table}[h]
\centering
\caption{FDR dip vs.\ plateau comparison for PPKTP at
$L = 3$\,mm, $w_0 = 500\,\mu$m, $\lambda_p = 405$\,nm.
$\sigma_{\rm narrow}$ is the conditional position width at the narrowest scanned filter; $\sigma_{\rm dip}$ is its minimum over the full $\delta\lambda_{\rm filter}\in[0.1, 200]$\,nm scan.
The degenerate (810 nm) row is bolded in each crystal type and is
the deepest dip in both cases.}
\label{tab:fdr_reduction}
\setlength{\tabcolsep}{4pt} 
\begin{tabular}{lccccc}
\hline
 & wavelength & $\sigma_{\rm narrow}$ & $\sigma_{\rm dip}$ & $\delta\lambda^{\rm dip}$ & reduction \\ 
  & (nm) & ($\mu$m) & ($\mu$m) & (nm) & (\%) \\
\hline
Type-I  & 700        & 9.47 & 8.63 &  4.99 &  8.92 \\
        & 750        & 9.38 & 8.61 &  9.56 &  8.20 \\
        & 780        & 9.36 & 8.80 & 14.77 &  5.94 \\
        & \textbf{810 (deg)} & \textbf{9.35} & \textbf{8.22} & \textbf{47.70} & \textbf{12.08} \\
        & 880        & 9.38 & 8.59 & 14.77 &  8.45 \\
        & 961        & 9.47 & 8.61 &  9.56 &  9.12 \\
\hline
Type-0  & 700        & 9.24 & 8.40 &  3.23 &  9.09 \\
        & 750        & 9.16 & 8.38 &  7.70 &  8.45 \\
        & 780        & 9.13 & 8.54 & 11.88 &  6.50 \\
        & \textbf{810 (deg)} & \textbf{9.13} & \textbf{8.01} & \textbf{40.66} & \textbf{12.20} \\
        & 880        & 9.16 & 8.36 &  9.56 &  8.67 \\
        & 961        & 9.24 & 8.39 &  6.19 &  9.25 \\
\hline
\end{tabular}
\end{table}

The most noticeable feature of Fig.~\ref{fig:fdr_ppktp} is the appearance of a clear FDR dip in the degenerate configuration $\lambda_s = 810$\,nm in both PPKTP Type-0 (dip at $\delta\lambda_{\rm filter}\approx 40.66$\,nm, $12\%$ reduction) and Type-I (dip at $\approx 47.70$\,nm, $12\%$ reduction). Interestingly, in degenerate operation, the location of the dip is exactly the phasematching bandwidth. Prior work has shown that nondegenerate FDR dip location found following the relation $\delta\lambda_{dip} = 1.35 \delta\lambda_{PM}$ \cite{Kuniyil2026entanglement}. This holds true in our analysis, in which two nondegenerate dips follws exactly this trend, see Fig.\ref{fig:fdr_ppktp}. In the BBO case, the FDR dip at degeneracy
is far in the filter width. The CPM crystal GVM coefficient is small (BBO uses two ordinary down converted photons, whose group-index curvature is the weakest of the three crystal classes considered here), and the on-axis phase matching peaks for the signal and idler emission lobes are correspondingly broad. At $L = 3$\,mm we find $\delta\lambda_{\rm PM}^{\rm deg, \,BBO} \approx 80$\,nm, this is outside the useful experimental operating range of any practical SPDC source. In the QPM crystals, with their stronger GVM, have narrower phase-matching bandwidths (Table~\ref{tab:dlPM_three_crystals}).

Physically, the wide degenerate bandwidth is the lobe-merging limit of the X-entanglement picture introduced in Ref.~\cite{gatti2009Xentanglement}.  Away from degeneracy, the on-axis $\mathrm{sinc}^{2}$ phase-matching function has two distinct peaks --- one at the signal wavelength and one at the idler --- separated by $|\lambda_i - \lambda_s|$.  Each peak has the GVM-limited width $\delta\lambda_{\rm PM}^{\rm GVM}$ derived in
Appendix~\ref{app:dlPM_derivation}.  As $\lambda_s \to 2\lambda_p$ the two peaks approach each other and merge into a single peak whose full width is the sum of the two contributions; the effective degenerate phase-matching bandwidth therefore inflates on approach to degeneracy.

This understanding force us to drop Eq.~\eqref{eq:dlPM_signal} for degenerate bandwidth: as $\lambda_s \to \lambda_i$, the group- index difference $|n_g(\lambda_s) - n_g(\lambda_i)|$ in the denominator goes to zero leading to mathematically undefined condition.  At this point, leading-order Taylor expansion on which Eq.~\eqref{eq:dlPM_signal} is based breaks down precisely because the GVM coefficient vanishes at degeneracy.  The next non-zero term in the expansion is the second-order group-velocity dispersion (GVD), and the appropriate closed form at exact degeneracy is Eq.~\eqref{eq:app_dlPM_GVD} of Appendix~\ref{app:GVD},
\begin{equation}
\delta\lambda_{\rm PM}^{\rm GVD}(L) =
\lambda_{\rm deg}\sqrt{\frac{5.566}{\pi\,|n_g'(\lambda_{\rm deg})|\,L}}\,,
\label{eq:dlPM_GVD_recall}
\end{equation}
which gives a finite answer set by the wavelength derivative of the group index at the degenerate wavelength, rather than the group-
index difference between nondegenerate signal and idler. For the three crystals at $L=3$\,mm, Eq.~\eqref{eq:dlPM_GVD_recall} yields $\delta\lambda_{\rm PM}^{\rm GVD} = 40.66$\,nm (PPKTP Type-0), $47.70$\,nm (PPKTP Type-I), and $78.3$\,nm (BBO Type-I).

\begin{figure}[h]
\centering
\includegraphics[width=\linewidth]{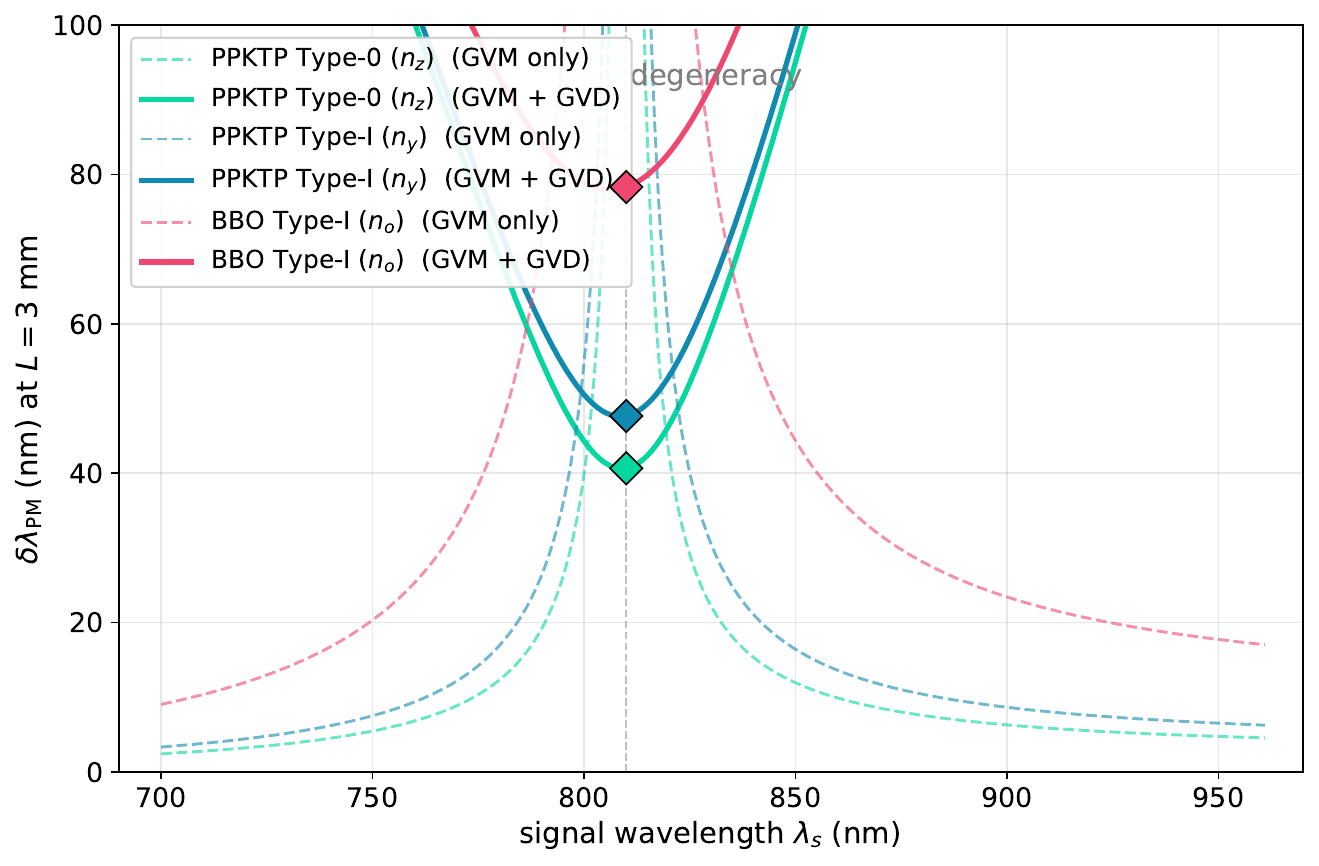}
\caption{Phase-matching bandwidth versus signal wavelength at $L = 3$\,mm, $\lambda_p = 405$\,nm.  Dashed curves: leading-order GVM expression Eq.~\eqref{eq:dlPM_signal}, which diverges as $\lambda_s$ approaches the degenerate point $2\lambda_p = 810$\,nm because $|n_g(\lambda_s) - n_g(\lambda_i)| \to 0$.  Diamond markers: leading-order GVD-only value Eq.~\eqref{eq:dlPM_GVD_recall} at
exact degeneracy.  Solid curves: full second-order expansion [Eq.~\eqref{eq:dlPM_GVD_recall}] that keeps both the linear GVM and quadratic GVD terms and is finite at all $\lambda_s$.  The full expression smoothly bridges the two limiting regimes and at exact degeneracy reproduces the GVD-only marker to better than $0.05$\,nm.}
\label{fig:bandwidth_degeneracy}
\end{figure}

Figure~\ref{fig:bandwidth_degeneracy} shows the resulting picture for the three crystals.  The GVM formula correctly captures the nondegenerate bandwidth and gives the practical operating point in the GVM-dominated regime; the GVD formula takes over in the narrow window around exact degeneracy and provides the finite value the GVM formula cannot.

The two closed forms are the leading-order results of the same Taylor expansion in two complementary limits. Equation~\eqref{eq:dlPM_signal} keeps only the first-order term in $\Omega = \omega_s - \omega_s^{(0)}$ and is exact when the linear GVM contribution dominates; Eq.~\eqref{eq:dlPM_GVD_recall} keeps the second-order term and is exact when the linear contribution vanishes by symmetry at $\lambda_s = \lambda_i = 2\lambda_p$.  Concretely, the cubic correction to the GVM term sets the crossover non-degeneracy $\varepsilon_{\rm GVD}$ ($\epsilon = (\lambda_i-\lambda_s)/(\lambda_i+\lambda_s)$) at which the two regimes match; below $\varepsilon_{\rm GVD}$ the GVD formula gives the right answer and above it the GVM formula does.  In the GVM-dominated regime, the bandwidth scales as $1/(\varepsilon L)$, while in the GVD-dominated regime it scales as $1/\sqrt{L}$ --- a direct signature of the second-order Taylor expansion at the degeneracy point.

The degenerate FDR dip is therefore not absent in any crystal class; it is a real and universal feature of the SPDC near-field FDR
profile, present whenever the spectral filter can be tuned to $\delta\lambda_{\rm filter} \approx \delta\lambda_{\rm PM}^{\rm deg}$.
CPM crystals push this filter requirement to impractically wide values; QPM crystals with their stronger GVM and narrower phase-matching windows bring it down to experimentally accessible widths. This is a qualitatively new feature of QPM-crystal FDR that goes beyond the CPM crystal phenomenology.  The agreement between the analytical $\delta\lambda_{\rm PM}^{\rm GVD}$ of $40.66$ and $47.70$\,nm and the observed PPKTP dip locations are within the $<1\%$ error (Table~\ref{tab:fdr_reduction}).
\section{discussion and outlook}
\label{sec:dscussion}
\subsection{Axis based anisotropy in far field regime for birefringence sensing}
The far-field system is characterized by axis-dependent conditional uncertainty: the walk-off axis has a narrower conditional uncertainty than its orthogonal, non–walk-off axis ($\sigma_{q_i\mid q_s = 0}^y< \sigma_{q_i\mid q_s = 0}^x$). This defines an anisotropy ellipse whose orientation is set by the crystal optic axis and controlled by the crystal length and spectral filter bandwidth, as analyzed in Sec. \ref{sec:crystal_length_effect}. The resulting imaging system has axis-dependent resolution—finer along the walk-off axis—which provides a calibrated spatial reference.

Such a source is naturally suited for birefringence sensing that uses spatial rather than polarization degrees of freedom. It is the spatial analogue of polarization-based quantum birefringence imaging, where retardance and principal-axis angle are obtained from polarization entanglement \cite{zhang2024quantum}. Here, however, the birefringence signature appears in the second-order spatial correlation without altering the polarization state. A birefringent object with principal axes misaligned from the source polarization imposes its own direction-dependent response on the conditional-correlation ellipse. The object’s principal-axis orientation is inferred from the rotation of the measured ellipse relative to the calibrated source frame, and the magnitude of its anisotropy from the change in ellipse eccentricity—yielding two continuous observables from a single coincidence measurement.

Quantification is enabled by the static, known source anisotropy: the ellipse orientation and eccentricity are first calibrated with no object present, so any subsequent rotation or eccentricity change can be attributed to the sample. A full separation of source and sample contributions, and a precise sensitivity analysis, require detailed modeling and are left for future work. Nonetheless, the demonstrated sensitivity of the transverse correlation width to the birefringence axis suggests that engineered SPDC anisotropy can provide an additional, spatially encoded contrast channel for birefringence and orientation sensing.

This kind of correlation strength anisotropy of the SPDC two-photon state along and across the walk-off plane was previously observed by Fedorov et al~\cite{fedorov2007anisotropy}, they used a Schmidt-decomposition analysis to show that the degree of entanglement (Schmidt number $K$) of the unconditioned state differs between the two transverse axes~\cite{law2004analysis}. Two features distinguish the present work from this. First, the controlled quantity is different: Fedorov \emph{et al.} characterized the Schmidt rank of the unconditioned two-photon state, whereas our observable is the conditional second-order moment $\sigma_{q_s|q_i}$ that directly sets the point-spread function of an imaging system. Second, the control mechanism is fundamentally different in kind. The control parameter in the Fedorov analysis is the angular acceptance: a spatial filter applied directly to the emission cone, which truncates the transverse-momentum distribution itself. Our control combines the crystal length with the spectral configuration, neither of which acts directly on the transverse momentum; both act only through the frequency dependence of the phase-matching function. These operations are not interchangeable, because the SPDC biphoton amplitude is non-separable in transverse momentum and frequency --- the X-shaped spatiotemporal structure of the two-photon correlation~\cite{gatti2009Xentanglement}. A spatial filter cuts the transverse-momentum field of this structure directly; a spectral selection reshapes the surviving transverse-momentum content only indirectly, through the asymmetric coupling between the spatial and spectral degrees of freedom encoded in the phase mismatch, where frequency enters through the group-velocity mismatch (GVM- linear term) and group-velocity dispersion (GVD, quadratic) while transverse momentum appears through diffraction and, along the walk-off axis alone, through the linear walk-off term. In Fedorov's picture the walk-off anisotropy is a static, intrinsic property of
the unconditioned state; in this investigation it becomes a quantity that can be switched on and off through a combination of geometric and spectral degrees of freedom.

A direct, experimentally accessible signature confirms that the control is genuinely spectral rather than an effective angular acceptance: the spectral filter's leverage over the correlation vanishes precisely at degeneracy. As the signal approaches the degenerate wavelength $\lambda_s\!\to\!2\lambda_p$, the group-velocity mismatch $|n_g(\lambda_s)-n_g(\lambda_i)|$ vanishes, the phase-matching bandwidth $\delta\lambda_{\mathrm{PM}}$ diverges, $R\!\to\!0$ and $\gamma\!\to\!1$; the walk-off narrowing then locks to its maximal filter-independent value, and no choice of bandwidth can suppress it. A spatial filter acting directly on the emission angle has no such a turning point in degeneracy. The appearance of this turning point at the group-velocity-matched wavelength is therefore a falsifiable fingerprint of a spectral-degree-of-freedom control --- one that the angular-acceptance mechanism structurally cannot reproduce.
\subsection{Implication: improved spatial resolution at degeneracy in QPM imaging}
\label{ssec:resolution_implication}

The observation that the degenerate FDR dip is deepest of all six configurations in both PPKTP types ($\approx 12\%$ in Table~\ref{tab:fdr_reduction}, compared with $5$--$9\%$ for the nondegenerate cases) has a direct practical consequence for biphoton imaging.  In a QPM based imaging source operated at the degenerate configuration, choosing the signal-arm filter bandwidth near the dip --- $41$ or $48$ \,nm (depending upon the type of crystal) seen in Fig.~\ref{fig:fdr_ppktp} --- reduces the conditional position uncertainty $\sigma_{x_s|x_i=0}$ to $12\%$ in both type-0 and type-I.

For an imaging scheme whose resolution is set by $\sigma_{x_s|x_i=0}$ (centroid measurement based super resolution \cite{toninelli2019resolution}, ghost-imaging point-spread function \cite{moreau2018resolution}, demonstration of super resolution at Heisenberg limit \cite{unternahrer2018super}, spatial resolution in quantum imaging with undetected photons \cite{fuenzalida2022resolution}), the $12\%$ tightening of $\sigma_x$ at the degenerate dip directly translates into a corresponding tightening of the imaging point-spread function. For an SPDC source designed with the flexibility to operate either degenerately or nondegenerately, the QPM degenerate configuration with a $40$--$50$\,nm filter therefore delivers a tighter conditional spatial
correlation than any other configuration in our scan.  This is a real advantage that has the potential to improve the biphoton based quantum imaging scheme by $12\%$.

\subsection{Filter-arm choice at degeneracy and signal--idler distinguishability}
\label{ssec:arm_choice_degen}

The dip-position scaling Eq.~\eqref{eq:dip_scaling} that holds for nondegenerate pairs becomes trivial at exact degeneracy ---
$(\lambda_i/\lambda_s)^{2}\to 1$ --- because the signal and idler are spectrally indistinguishable.  Consequently, the signal-arm and idler-arm filter choices coincide at the degenerate point, and the dip is located at the same filter width regardless of which arm the filter is placed in.  As the operating point is moved away from degeneracy, this signal--idler symmetry is broken, and Eq.~\eqref{eq:dip_scaling} kicks in with its $(\lambda_i/\lambda_s)^{2}$ prefactor.  The smooth crossover from the symmetric degenerate point to the asymmetric nondegenerate regime is therefore a continuous function of the nondegeneracy.

\section{conclusion}
This work shows that the spectral filter is a direct design knob over the conditional correlation uncertainty of biphotons in both the momentum and the position domains, and that the choice of filter bandwidth directly influences conditional uncertainty parameters in the transverse direction. In the conditional momentum domain, the walk-off axis of critically phase-matched nonlinear crystals provides a width-narrowing advantage with crystal length, and our analytical closed form support the simulation results. In addition, we establish that the choice of which arm carries the spectral filter is itself a design parameter: filtering in the idler arm rather than the signal arm rescales the conditional bandwidth (and therefore the conditional uncertainty) by the kinematic factor $(\lambda_i/\lambda_s)^2$, a closed-form ratio that follows directly from the wavelength-to-frequency Jacobian and that holds independently of the crystal class. In the position domain, our treatment confirms that the FDR profile first observed in nondegenerate CPM crystals also appears in QPM-crystal sources, and we further prove that the FDR profile is universal and is present even at exact degeneracy. The apparent insensitivity of the conditional position width to filter bandwidth reported in the earlier BBO Type-I study is an artifact of the much wider phase-matching window of BBO Type-I, which compensates for the degenerate dip beyond the practically accessible filter range. However, in QPM crystals, with their narrower phase-matching windows, the degenerate dip is brought down to filter widths that are experimentally accessible. As a result, the operation of a QPM source at the degenerate point with a 40–50 nm filter improves the conditional position width by about 12\% over the baseline of the narrow-filter, the deepest improvement of any configuration we examined and the most directly useful operating point for QPM-based quantum imaging.

\section*{Acknowledgments}
The authors thank Yingwen Zhang for useful discussions and comments that improved this work.

\appendix
\section{Derivation of the phase-matching bandwidth $\delta\lambda_{\rm PM}$ from the Sellmeier}
\label{app:dlPM_derivation}
This section derives the closed-form expression for the GVM-limited and GVD-limited phase-matching bandwidths explained in  Sec.~\ref{ssec:idler_arm} and \ref{ssec:degenerate_dip_ppktp}. The result is obtained using the standard Sellmeier equations for crystals \cite{kato2002sellmeier, emanueli2003temperature}. The derivation follow the standard SPDC treatment (see Ref.~\cite{boyd2020nonlinear}), specialized to the on-axis emission and to the choice of which arm (signal or idler) carries the spectral filter.

\subsection{The on-axis phase mismatch}
\label{app:on_axis_phase}
Ideally, the signal and idler have a spectral distribution even though the pump is considered strictly single frequency laser emission. This distribution arises due to the relaxation of the strict phasematching efficiency driven by the Sinc function of SPDC emission. For that we can Taylor expand individual emission frequencies as
\begin{equation}
    k_s(\omega_s^{(0)}+\Omega) = k_s(\omega_s^{(0)}) + k_s^{'}\Omega+\frac{1}{2}k_s^{''}\Omega^2+.......
    \label{signal_typlor_expansion}
\end{equation}
where $k' = dk/d\omega = 1/v_g$, $v_g$ is the group velocity of the light. similarly, for idler
\begin{equation}
    k_i(\omega_i^{(0)}-\Omega) = k_i(\omega_i^{(0)}) - k_i^{'}\Omega+\frac{1}{2}k_i^{''}\Omega^2+.......
    \label{idler_typlor_expansion}
\end{equation}
where $\omega$ denotes the angular frequency, and $\Omega$ is the detuning frequency defined as $\Omega = \omega - \omega^{(0)}$. And $k''$ is the second-order derivative of the wavenumber with respect to the angular frequency $\omega$. Phase matching in SPDC can be defined as follows: 
\begin{equation}
    \Delta k = k_p - k_s(\omega_s^{(0)}+\Omega) - k_i(\omega_i^{0}-\Omega)
    \label{eq:phaseMatching}
\end{equation}
substituting Eqs.~\ref{signal_typlor_expansion} and \ref{idler_typlor_expansion} into equation Eq.~\eqref{eq:phaseMatching}, we get
\begin{align}
    \Delta k(\Omega) = \Delta k_z^{(0)} - G_m-\tau\Omega-\frac{1}{2}k_{eff}^{''}\Omega^2
    \label{eq:phasematching-frequency_domain}
\end{align}
where $\Delta k_z^{(0)}$ is the longitudinal phase matching term at a strictly central wavelength and therefore we can approximate $\Delta k_z^{(0)} - G_m= 0$, $G_m = 2\pi/\Lambda$,$|\Lambda = 2\pi/(k_p - k_s - k_i)|$ is specific to the QPM crystal that defines the noncritical phase matching in the QPM crystal,  $\tau$ is the GVM element in the equation equal to $\tau = (n_g(\lambda_i) - n_g(\lambda_s))/c$,    $k_{eff}^{''} = k_s^{''}+k_i^{''}= d^2k_s/d\omega_s^2+ d^2k_i/d\omega_i^2$
\subsection{Taylor expansion in the signal detuning}
\label{app:taylor}

Defining the signal detuning $\Omega = \omega_s - \omega_s^{(0)}$ and expanding Eq.~\eqref{eq:phasematching-frequency_domain} to leading order in $\Omega$, the first-order coefficient is
\begin{align}
\frac{\partial \Delta k_z}{\partial \omega_s}\bigg|_{\omega_s^{(0)}}
& = -\frac{\partial k_s}{\partial \omega_s} - \frac{\partial k_i}{\partial \omega_i}\,\frac{\partial \omega_i}{\partial \omega_s} \notag\\
& = -\frac{\partial k_s}{\partial \omega_s} + \frac{\partial k_i}{\partial \omega_i}
  = -\left(\frac{1}{v_g(\lambda_s)} - \frac{1}{v_g(\lambda_i)}\right),
\label{eq:app_GVM}
\end{align}
where we used energy conservation $\partial \omega_i / \partial \omega_s = -1$ and the definition of the group velocity $\partial k/\partial \omega = 1/v_g$.  Defining the group-velocity-mismatch coefficient
\begin{equation}
\tau_{\rm GVM}(\lambda_s) \;\equiv\; \frac{1}{v_g(\lambda_s)} - \frac{1}{v_g(\lambda_i)}\,,
\label{eq:app_tau_def}
\end{equation}
the linearised phase mismatch reads
\begin{equation}
\Delta k_z(\omega_s) \;\approx\; -\,\tau_{\rm GVM}\,\Omega\,.
\label{eq:app_Dkz_lin}
\end{equation}
The quantity $\tau_{\rm GVM}$ has units of inverse velocity --- it is the time delay per unit crystal length between the signal and idler wavepackets.
\subsection{Express $\tau_{\rm GVM}$ through the group index}
\label{app:group_index}

The group velocity is related to the group index by $v_g = c/n_g$, where the group index is computable directly from the refractive index by
\begin{equation}
n_g(\lambda) \;=\; n(\lambda) - \lambda\,\frac{dn(\lambda)}{d\lambda}\,.
\label{eq:app_ng_def}
\end{equation}
This identity is derived from the chain rule in $n_g = n + \omega(dn/d\omega)$ after substituting $\omega = 2\pi c/\lambda$. Combining Eqs.~\eqref{eq:app_tau_def} and~\eqref{eq:app_ng_def},
\begin{equation}
\tau_{\rm GVM}(\lambda_s) \;=\; \frac{n_g(\lambda_s) - n_g(\lambda_i)}{c}\,,
\label{eq:app_tau_ng}
\end{equation}
which is the key relation linking the framework to the Sellmeier: $n_g(\lambda)$ is computable by analytic differentiation of the Sellmeier of the relevant daughter polarisation, or by symmetric finite difference at small step size ($h = 10^{-4}\,\mu$m suffices to machine precision).
\subsection{The sinc$^{2}$ FWHM}
\label{app:sinc_FWHM}
The phase-matching weight in the biphoton amplitude is $\mathrm{sinc}^{2}(\Delta k_z\,L/2)$.  By direct solution of $\mathrm{sinc}^{2}(x) = 1/2$, the half-width of $\mathrm{sinc}^{2}$ in its argument is $x_{1/2} \approx 1.391$, so the full width at half maximum of $\Delta k_z$ is
\begin{equation}
\Delta k_z^{\rm FWHM} \;=\; \frac{2 \times 2 \times 1.391}{L} \;=\; \frac{5.566}{L}\,.
\label{eq:app_Dkz_FWHM}
\end{equation}
The numerical factor $5.566$ is a property of the sinc$^{2}$ function alone and is independent of the crystal, wavelength or geometry.  Substituting Eq.~\eqref{eq:app_Dkz_lin} into Eq.~\eqref{eq:app_Dkz_FWHM} gives the FWHM in signal angular frequency,
\begin{equation}
\delta\omega_{\rm PM} \;=\; \frac{5.566}{L\,|\tau_{\rm GVM}|}
\;=\; \frac{5.566\,c}{L\,|n_g(\lambda_s) - n_g(\lambda_i)|}\,.
\label{eq:app_domega_PM}
\end{equation}
We have verified Eq.~\eqref{eq:app_domega_PM} against the full angular-integrated SPDC emission spectrum simulated for BBO Type-I at $L = 10$\,mm, $\lambda_s = 780$\,nm, $\rho = 0$ (no walk-off): the analytical formula predicts $\delta\lambda_{\rm PM}^{(s)} = 13.71$\,nm and $\delta\lambda_{\rm PM}^{(i)} = 15.99$\,nm; the numerical FWHMs of the angular-integrated emission spectrum are $13.07$\,nm and $15.08$\,nm, respectively, with the residual ${\sim}5\%$ narrowing attributable to the integration over off-axis modes [Sec.~\ref{app:on_axis_phase}].  The ratio of the two peaks is $1.165$ analytically and $1.165$ numerically, in agreement to better than $0.1\%$ with the predicted $(\lambda_i/\lambda_s)^{2}$ scaling of Eq.~\eqref{eq:idler_scaling}

Equation~\eqref{eq:app_domega_PM} is symmetric under the signal--idler exchange $\lambda_s \leftrightarrow \lambda_i$, because $\tau_{\rm GVM}$ is itself a pair property of the biphoton rather than a single-photon property.  This symmetry is the analytical origin of the signal--idler scaling discussed below.

\subsection{Conversion to wavelength: signal-arm vs idler-arm filtering}
\label{app:wavelength_conversion}
The angular frequency and wavelength are related by $\omega = 2\pi c/\lambda$ which gives
\begin{equation}
|\delta\omega| \;=\; \frac{2\pi c}{\lambda^{2}}\,|\delta\lambda|\,,
\label{eq:app_omega_lambda}
\end{equation}
where the wavelength $\lambda$ in the prefactor is the wavelength of the photon being filtered.  When the spectral filter sits on the signal arm, $\lambda = \lambda_s$; when it sits on the idler arm, $\lambda = \lambda_i$.  Applying Eq.~\eqref{eq:app_omega_lambda} to Eq.~\eqref{eq:app_domega_PM} gives the two closed-form bandwidths
\begin{align}
\delta\lambda_{\rm PM}^{(s)}(\lambda_s,L)
\;=\;
\frac{5.566\,\lambda_s^{\,2}}{2\pi\,L\,|n_g(\lambda_s) - n_g(\lambda_i)|}\,,\\
\qquad
\delta\lambda_{\rm PM}^{(i)}(\lambda_i,L)
\;=\;
\frac{5.566\,\lambda_i^{\,2}}{2\pi\,L\,|n_g(\lambda_s) - n_g(\lambda_i)|}
\label{eq:app_dlPM_arms}
\end{align}
The denominators are identical because they encode the GVM coefficient of the biphoton pair, not a single-photon property.  The ratio is
therefore exactly the kinematic factor
\begin{equation}
\frac{\delta\lambda_{\rm PM}^{(i)}}{\delta\lambda_{\rm PM}^{(s)}} \;=\; \left(\frac{\lambda_i}{\lambda_s}\right)^{\!2}\,,
\label{eq:app_arm_ratio}
\end{equation}
recovering the signal--idler scaling of
Eqs.~\eqref{eq:idler_scaling} and~\eqref{eq:dip_scaling} of the
main text.  Equation~\eqref{eq:app_arm_ratio} is an analytical
identity --- not an empirical fit --- because any higher-order
corrections to the absolute bandwidths affect both arms by the
same multiplicative factor and cancel in the ratio.
\subsection{Degenerate regime: the second-order Taylor expansion}
\label{app:GVD}

In exact degeneracy $\lambda_s = \lambda_i = 2\lambda_p$ the GVM
coefficient vanishes by symmetry ($\tau_{\rm GVM}\to 0$), and the
linearized expansion of Sec.~\ref{app:taylor} breaks down.  The
next non-zero term in the Taylor expansion of
Eq.~\eqref{eq:phasematching-frequency_domain} is second-order in $\Omega$:
\begin{equation}
\Delta k_z(\omega_s) \;\approx\; -\tfrac{1}{2}\,k''_{\rm eff}\,\Omega^{2},
\qquad
k''_{\rm eff} \;=\; \frac{\lambda_{\rm deg}^{\,2}}{\pi c^{2}}\,|n_g'(\lambda_{\rm deg})|\,,
\label{eq:app_Dkz_GVD}
\end{equation}
where $n_g'(\lambda) \equiv dn_g/d\lambda$ is the wavelength
derivative of the group index, obtained by a second differentiation
of the Sellmeier.  The sinc$^{2}$ FWHM condition
Eq.~\eqref{eq:app_Dkz_FWHM} now reads
$|k''_{\rm eff}\,\Omega^{2}\,L/4| = 1.391$, yielding
\begin{equation}
\Omega_{\rm FWHM} \;=\; 2\sqrt{\frac{2.782}{L\,|k''_{\rm eff}|}}\,.
\label{eq:app_omega_FWHM_GVD}
\end{equation}
Converting to wavelength at $\lambda = \lambda_{\rm deg}$:
\begin{equation}
\delta\lambda_{\rm PM}^{\rm GVD}(L)
\;=\;
\lambda_{\rm deg}\sqrt{\frac{5.566}{\pi\,|n_g'(\lambda_{\rm deg})|\,L}}\,.
\label{eq:app_dlPM_GVD}
\end{equation}
At degeneracy $\lambda_s = \lambda_i$ so the signal-arm and idler-arm
conversions of Eq.~\eqref{eq:app_omega_lambda} coincide and
Eq.~\eqref{eq:app_arm_ratio} reduces to unity.  The $L^{-1/2}$
scaling of Eq.~\eqref{eq:app_dlPM_GVD} contrasts with the $L^{-1}$
scaling of Eq.~\eqref{eq:app_dlPM_arms} and is a direct signature
of the second-order Taylor expansion: at the degeneracy point the
sinc$^{2}$ width scales with the square root of $L\,|k''_{\rm eff}|$
rather than its inverse.

\subsection{Application to the four crystal classes}
\label{app:four_crystals}

The structure of the derivation is identical for every crystal
class considered in this work; the only quantity that changes
between classes is which Sellmeier function enters the group
index of the signal and idler polarisations:
\begin{itemize}
\item
\emph{PPKTP Type-0} ($z\to z+z$): both daughters use $n_z(\lambda)$,
so $n_g^{(s)} = n_g^{(i)} = n_g[n_z(\lambda)]$.
\item
\emph{PPKTP Type-I} ($z\to y+y$): both daughters use $n_y(\lambda)$,
so $n_g^{(s)} = n_g^{(i)} = n_g[n_y(\lambda)]$.
\item
\emph{BBO Type-I} ($e\to o+o$): both daughters ordinary, so
$n_g^{(s)} = n_g^{(i)} = n_g[n_o(\lambda)]$.
\end{itemize}
In all four cases the same closed form
Eqs.~\eqref{eq:app_dlPM_arms}--\eqref{eq:app_dlPM_GVD} applies; the
crystal class only determines which Sellmeier is differentiated.
The BBO Type-II case is qualitatively different from the other
three because the cross-polarised signal--idler pair produces a
non-zero GVM \emph{at every wavelength including degeneracy},
making the Type-II emission spectrum nearly wavelength-independent
(in contrast to the strongly $\lambda_s$-dependent bandwidth of
the three co-polarised cases).
\begin{table*}
\centering
\caption{Phase-matching bandwidth $\delta\lambda_{\rm PM}(\lambda_s, L)$
 in nm on the signal arm, computed analytically from the Sellmeier
 via Eq.~\eqref{eq:app_dlPM_arms} (nondegenerate) and
 Eq.~\eqref{eq:app_dlPM_GVD} (degenerate $\lambda_s = 2\lambda_p = 810$\,nm,
 marked with an asterisk). Pump $\lambda_p = 405$\,nm, $T = 25^\circ$C. The idler-arm
 values follow by multiplying each nondegenerate entry by
 $(\lambda_i/\lambda_s)^{2}$ [Eq.~\eqref{eq:app_arm_ratio}]; the degenerate
 column is unchanged since $\lambda_s = \lambda_i$. An experimentalist
 reads off $R = \delta\lambda_{\rm filter}/\delta\lambda_{\rm PM}$
 directly from this table and substitutes into
 Eq.~\eqref{eq:sigma_walkoff}.}
\label{tab:dlPM_three_crystals}
\begin{tabular}{lc|cccccccccc}
\hline\hline
crystal & $L$ (mm) & 700 & 730 & 750 & 780 & 810 & 830 & 850 & 880 & 910 & 961 \\
        &          & nm  & nm  & nm  & nm  & nm  & nm  & nm  & nm  & nm  & nm  \\
\hline
\multirow{4}{*}{PPKTP Type-0 ($n_z$)}
 & 1 &  7.24 & 11.32 & 16.39 & 36.91 & $70.4^{*}$ &  66.72 &  35.82 & 22.69 & 17.54 & 13.64 \\
 & 3 &  2.41 &  3.77 &  5.46 & 12.30 & $40.66^{*}$ &  22.24 &  11.94 &  7.56 &  5.85 &  4.55 \\
 & 5 &  1.45 &  2.26 &  3.28 &  7.38 & $31.5^{*}$ &  13.34 &   7.16 &  4.54 &  3.51 &  2.73 \\
 & 9 &  0.80 &  1.26 &  1.82 &  4.10 & $23.5^{*}$ &   7.41 &   3.98 &  2.52 &  1.95 &  1.52 \\
\hline
\multirow{4}{*}{PPKTP Type-I ($n_y$)}
 & 1 &  9.96 & 15.56 & 22.53 & 50.74 & $82.6^{*}$ &  91.72 &  49.24 & 31.20 & 24.12 & 18.76 \\
 & 3 &  3.32 &  5.19 &  7.51 & 16.91 & $47.70^{*}$ &  30.57 &  16.41 & 10.40 &  8.04 &  6.25 \\
 & 5 &  1.99 &  3.11 &  4.51 & 10.15 & $36.9^{*}$ &  18.34 &   9.85 &  6.24 &  4.82 &  3.75 \\
 & 9 &  1.11 &  1.73 &  2.50 &  5.64 & $27.5^{*}$ &  10.19 &   5.47 &  3.47 &  2.68 &  2.08 \\
\hline
\multirow{4}{*}{BBO Type-I ($e\to o+o$, $n_o$)}
 & 1 & 27.10 & 42.19 & 60.97 & 137.09 & $135.7^{*}$ & 247.75 & 133.07 & 84.42 & 65.39 & 51.08 \\
 & 3 &  9.03 & 14.06 & 20.32 &  45.70 & $78.3^{*}$  &  82.58 &  44.36 & 28.14 & 21.80 & 17.03 \\
 & 5 &  5.42 &  8.44 & 12.19 &  27.42 & $60.7^{*}$  &  49.55 &  26.61 & 16.88 & 13.08 & 10.22 \\
 & 9 &  3.01 &  4.69 &  6.77 &  15.23 & $45.2^{*}$  &  27.53 &  14.79 &  9.38 &  7.27 &  5.68 \\
\hline\hline
\end{tabular}
\medskip
\noindent
{$^{*}$degenerate ($\lambda_s = \lambda_i = 810$\,nm),
computed from the GVD-limited expression
Eq.~\eqref{eq:app_dlPM_GVD} with $L^{-1/2}$ scaling, rather than the
nondegenerate $L^{-1}$ scaling of Eq.~\eqref{eq:app_dlPM_arms}}.
\end{table*}

\subsection{Underlying approximations}
\label{app:assumptions}

The derivation rests on four well-controlled approximations.

\paragraph*{(a)~Paraxial approximation in $\Delta k_z$.}
The on-axis evaluation $q_s = q_i = 0$ used in
Eq.~\eqref{eq:near_field_phasematching} is the paraxial leading order; off-axis
transverse momenta contribute terms of order $q^{2}/k$ that are
small whenever the emission cone is much narrower than the
crystal numerical aperture.  For the typical
$w_0 = 500\,\mu$m / $L \le 10$\,mm parameters of this work these
corrections are at the $\sim 5\%$ level (see
Sec.~\ref{app:on_axis_phase}).

\paragraph*{(b)~Linear GVM expansion.}
The truncation of the Taylor expansion at first order in $\Omega$
[Eq.~\eqref{eq:app_Dkz_lin}] is valid when the signal sits well
away from degeneracy.  The breakdown at the degenerate point is
handled by the second-order expansion of Sec.~\ref{app:GVD},
which is itself valid until the cubic corrections become important
(i.e.\ for $\delta\lambda_{\rm filter} \ll \lambda_{\rm deg}$).

\paragraph*{(c)~Monochromatic CW pump.}
Energy conservation $\omega_p = \omega_s + \omega_i$ has been used
with a fixed pump frequency.  For pulsed pumping with bandwidth
$\delta\omega_p$, an additional convolution with the pump spectrum
enters and effectively replaces $\tau_{\rm GVM}$ by an averaged
mismatch.  The CW assumption is valid for the standard SPDC
imaging sources considered in this work.

\paragraph*{(d)~On-axis emission detection.}
The closed-form Eqs.~\eqref{eq:app_dlPM_arms} and~\eqref{eq:app_dlPM_GVD} give the FWHM of the sinc$^{2}$ at $q_s = q_i = 0$. Therefore, conditional correlation is measured at the center at the traverse point x = y = 0 with point -to-point correlation analysis Eq.~\eqref{eq:Delta-q}. 
\bibliography{ref} 

@article{howell2004realization,
  title={Realization of the Einstein-Podolsky-Rosen paradox using momentum-and position-entangled photons from spontaneous parametric down conversion},
  author={Howell, John C and Bennink, Ryan S and Bentley, Sean J and Boyd, Robert W},
  journal={Physical Review Letters},
  volume={92},
  number={21},
  pages={210403},
  year={2004},
  publisher={APS},
  doi = {10.1103/PhysRevLett.92.210403}
}

@article{walborn2010spatial,
  title={Spatial correlations in parametric down-conversion},
  author={Walborn, Stephen P and Monken, CH and P{\'a}dua, S and Ribeiro, PH Souto},
  journal={Physics Reports},
  volume={495},
  number={4-5},
  pages={87--139},
  year={2010},
  publisher={Elsevier},
  doi     = {10.1016/j.physrep.2010.06.003}
}

@article{kuniyil2021efficient,
  title={Efficient coupling of down-converted photon pairs into single mode fiber},
  author={Kuniyil, Hashir and Durak, Kadir},
  journal={Optics Communications},
  volume={493},
  pages={127038},
  year={2021},
  publisher={Elsevier},
  doi ={10.1016/j.optcom.2021.127038}
}

@article{pearce2026quantum,
  title={Quantum imaging with correlated photon pairs},
  author={Pearce, Emma and Nothlawala, Fazilah and Forbes, Andrew and Padgett, Miles J},
  journal={Nature Reviews Methods Primers},
  volume={6},
  number={1},
  pages={17},
  year={2026},
  publisher={Nature Publishing Group UK London},
  doi = {10.1038/s43586-025-00468-x}
}

@article{law2004analysis,
  title={Analysis and interpretation of high transverse entanglement in optical parametric down conversion},
  author={Law, CK and Eberly, JH},
  journal={Physical Review Letters},
  volume={92},
  number={12},
  pages={127903},
  year={2004},
  publisher={APS},
  doi = {10.1103/PhysRevLett.92.127903}
}

@article{toninelli2019resolution,
  title={Resolution-enhanced quantum imaging by centroid estimation of biphotons},
  author={Toninelli, Ermes and Moreau, Paul-Antoine and Gregory, Thomas and Mihalyi, Adam and Edgar, Matthew and Radwell, Neal and Padgett, Miles},
  journal={Optica},
  volume={6},
  number={3},
  pages={347--353},
  year={2019},
  publisher={Optical Society of America},
  doi = {10.1364/OPTICA.6.000347}
}

@article{gregory2021noise,
  title={Noise rejection through an improved quantum illumination protocol},
  author={Gregory, Thomas and Moreau, P-A and Mekhail, Simon and Wolley, Osian and Padgett, MJ},
  journal={Scientific reports},
  volume={11},
  number={1},
  pages={21841},
  year={2021},
  publisher={Nature Publishing Group UK London},
  doi = {10.1038/s41598-021-01122-8}
}

@article{kuniyil2022noise,
  title={Noise-tolerant object detection and ranging using quantum correlations},
  author={Kuniyil, Hashir and Ozel, Helin and Yilmaz, Hasan and Durak, Kadir},
  journal={Journal of Optics},
  volume={24},
  number={10},
  pages={105201},
  year={2022},
  publisher={IOP Publishing},
  doi = {10.1088/2040-8986/ac899c}
}

@article{jedrkiewicz2004detection,
  title={Detection of sub-shot-noise spatial correlation in high-gain parametric down conversion},
  author={Jedrkiewicz, O and Jiang, Y-K and Brambilla, E and Gatti, A and Bache, M and Lugiato, LA and Di Trapani, P},
  journal={Physical Review Letters},
  volume={93},
  number={24},
  pages={243601},
  year={2004},
  publisher={APS},
  doi = {10.1103/PhysRevLett.93.243601}
}

@article{brida2010experimental,
  title={Experimental realization of sub-shot-noise quantum imaging},
  author={Brida, Giorgio and Genovese, Marco and Ruo Berchera, Ivano},
  journal={Nature Photonics},
  volume={4},
  number={4},
  pages={227--230},
  year={2010},
  publisher={Nature Publishing Group},
  doi = {10.1038/nphoton.2010.29}
}

@article{pittman1996two,
  title={Two-photon geometric optics},
  author={Pittman, TB and Strekalov, DV and Klyshko, DN and Rubin, MH and Sergienko, AV and Shih, YH},
  journal={Physical Review A},
  volume={53},
  number={4},
  pages={2804},
  year={1996},
  publisher={APS},
  doi = {10.1103/PhysRevA.53.2804}
}

@article{lemos2014quantum,
  title={Quantum imaging with undetected photons},
  author={Lemos, Gabriela Barreto and Borish, Victoria and Cole, Garrett D and Ramelow, Sven and Lapkiewicz, Radek and Zeilinger, Anton},
  journal={Nature},
  volume={512},
  number={7515},
  pages={409--412},
  year={2014},
  publisher={Nature Publishing Group},
  doi = {10.1038/nature13586}
}

@article{lahiri2015theory,
  title={Theory of quantum imaging with undetected photons},
  author={Lahiri, Mayukh and Lapkiewicz, Radek and Lemos, Gabriela Barreto and Zeilinger, Anton},
  journal={Physical Review A},
  volume={92},
  number={1},
  pages={013832},
  year={2015},
  publisher={APS},
  doi = {10.1103/PhysRevA.92.013832}
}

@article{cameron2024adaptive,
  title={Adaptive optical imaging with entangled photons},
  author={Cameron, Patrick and Courme, Baptiste and Verni{\`e}re, Chlo{\'e} and Pandya, Raj and Faccio, Daniele and Defienne, Hugo},
  journal={Science},
  volume={383},
  number={6687},
  pages={1142--1148},
  year={2024},
  publisher={American Association for the Advancement of Science},
  doi = {10.1126/science.adk782}
}

@article{zheng2025position,
  title={Position-correlated biphoton wavefront sensing for quantum adaptive imaging},
  author={Zheng, Yi and Liu, Zhao-Di and Tang, Jian-Shun and Xu, Jin-Shi and Li, Chuan-Feng and Guo, Guang-Can},
  journal={Light: Science \& Applications},
  volume={14},
  number={1},
  pages={311},
  year={2025},
  publisher={Nature Publishing Group UK London},
  doi = {10.1038/s41377-025-02024-4}
}

@article{verniere2026entanglement,
  title={Entanglement-enabled image transmission through complex media},
  author={Verni{\`e}re, Chlo{\'e} and Guitter, Rapha{\"e}l and Courme, Baptiste and Defienne, Hugo},
  journal={Nature Physics},
  pages={1--8},
  year={2026},
  publisher={Nature Publishing Group UK London},
  doi = {10.1038/s41567-026-03265-9}
}

@article{aspden2013epr,
  title={EPR-based ghost imaging using a single-photon-sensitive camera},
  author={Aspden, Reuben S and Tasca, Daniel S and Boyd, Robert W and Padgett, Miles J},
  journal={New Journal of Physics},
  volume={15},
  number={7},
  pages={073032},
  year={2013},
  publisher={IOP Publishing},
  doi = {10.1088/1367-2630/15/7/073032}
}

@article{moreau2018ghost,
  title={Ghost imaging using optical correlations},
  author={Moreau, Paul-Antoine and Toninelli, Ermes and Gregory, Thomas and Padgett, Miles J},
  journal={Laser \& Photonics Reviews},
  volume={12},
  number={1},
  pages={1700143},
  year={2018},
  publisher={Wiley Online Library},
  doi = {10.1002/lpor.201700143}
}

@article{joobeur1996coherence,
  title     = {Coherence properties of entangled light beams generated by
               parametric down-conversion: Theory and experiment},
  author    = {Joobeur, Adel and Saleh, Bahaa E. A. and Larchuk, Timothy S.
               and Teich, Malvin C.},
  journal   = {Phys. Rev. A},
  volume    = {53},
  pages     = {4360--4371},
  year      = {1996},
  publisher = {American Physical Society},
  doi       = {10.1103/PhysRevA.53.4360}
}

@article{moreau2018resolution,
  title={Resolution limits of quantum ghost imaging},
  author={Moreau, Paul-Antoine and Toninelli, Ermes and Morris, Peter A and Aspden, Reuben S and Gregory, Thomas and Spalding, Gabriel and Boyd, Robert W and Padgett, Miles J},
  journal={Optics express},
  volume={26},
  number={6},
  pages={7528--7536},
  year={2018},
  publisher={Optical Society of America},
  doi = {10.1364/OE.26.007528}
}

@article{fuenzalida2022resolution,
  title={Resolution of quantum imaging with undetected photons},
  author={Fuenzalida, Jorge and Hochrainer, Armin and Lemos, Gabriela Barreto and Ortega, Evelyn A and Lapkiewicz, Radek and Lahiri, Mayukh and Zeilinger, Anton},
  journal={Quantum},
  volume={6},
  pages={646},
  year={2022},
  publisher={Verein zur F{\"o}rderung des Open Access Publizierens in den Quantenwissenschaften},
  doi = {10.22331/q-2022-02-09-646}
}

@article{pires2011type,
  title={Type-I spontaneous parametric down-conversion with a strongly focused pump},
  author={Pires, H Di Lorenzo and Coppens, FMGJ and Van Exter, MP},
  journal={Physical Review A},
  volume={83},
  number={3},
  pages={033837},
  year={2011},
  publisher={APS},
  doi = {10.1103/PhysRevA.83.033837}
}

@article{Schneeloch2016TransverseSPDC,
  author    = {Schneeloch, J. and Howell, J. C.},
  title     = {Introduction to the transverse spatial correlations in spontaneous parametric down-conversion through the biphoton birth zone},
  journal   = {Journal of Optics},
  year      = {2016},
  volume    = {18},
  number    = {5},
  pages     = {053501},
  month     = may,
  doi       = {10.1088/2040-8978/18/5/053501},
  publisher = {IOP Publishing}
}

@article{kuniyil2026entanglement,
  title={Entanglement certification in bulk nonlinear crystals for degenerate and nondegenerate spontaneous parametric down conversion: Spectral filter effects on transverse spatial correlations},
  author={Kuniyil, Hashir and Ali, Asad and Al-Kuwari, Saif},
  journal={Physical Review A},
  volume={113},
  number={5},
  pages={052607},
  year={2026},
  publisher={APS},
  doi = {10.1103/drvp-dv4b}
}

@article{kato2002sellmeier,
  title={Sellmeier and thermo-optic dispersion formulas for KTP},
  author={Kato, Kiyoshi and Takaoka, Eiko},
  journal={Applied optics},
  volume={41},
  number={24},
  pages={5040--5044},
  year={2002},
  publisher={Optical Society of America},
  doi = {10.1364/AO.41.005040}
}

@article{emanueli2003temperature,
  title={Temperature-dependent dispersion equations for KTiOPO4 and KTiOAsO4},
  author={Emanueli, Shai and Arie, Ady},
  journal={Applied optics},
  volume={42},
  number={33},
  pages={6661--6665},
  year={2003},
  publisher={Optical Society of America}, 
  doi = {10.1364/AO.42.006661}
}

@article{brambilla2025,
  author    = {Brambilla, E. and Gatti, A. and Jedrkiewicz, O.},
  title     = {Certifying spatial entanglement between non-degenerate
               photon pairs with a camera},
  journal   = {Opt. Lett.},
  volume    = {50},
  pages     = {4854},
  year      = {2025},
  doi       = {10.1364/OL.553180}
}

@article{gatti2009xentanglement,
  title     = {$X$ Entanglement: The Nonfactorable Spatiotemporal Structure
               of Biphoton Correlation},
  author    = {Gatti, Alessandra and Brambilla, Enrico and Lugiato, Luigi A.},
  journal   = {Phys. Rev. Lett.},
  volume    = {102},
  pages     = {223601},
  year      = {2009},
  publisher = {American Physical Society},
  doi       = {10.1103/PhysRevLett.102.223601}
}

@article{zhang2024quantum,
  title={Quantum imaging of biological organisms through spatial and polarization entanglement},
  author={Zhang, Yide and He, Zhe and Tong, Xin and Garrett, David C and Cao, Rui and Wang, Lihong V},
  journal={Science Advances},
  volume={10},
  number={10},
  pages={eadk1495},
  year={2024},
  publisher={American Association for the Advancement of Science},
  doi = {10.1126/sciadv.adk1495}
}

@article{fedorov2007anisotropy,
  author    = {Fedorov, M. V. and Efremov, M. A. and Volkov, P. A.
               and Moreva, E. V. and Straupe, S. S. and Kulik, S. P.},
  title     = {Anisotropically and high entanglement of biphoton states
               generated in spontaneous parametric down-conversion},
  journal   = {Phys. Rev. Lett.},
  volume    = {99},
  pages     = {063901},
  year      = {2007},
  doi       = {10.1103/PhysRevLett.99.063901}
}

@article{unternahrer2018super,
  title={Super-resolution quantum imaging at the Heisenberg limit},
  author={Untern{\"a}hrer, Manuel and Bessire, B{\"a}nz and Gasparini, Leonardo and Perenzoni, Matteo and Stefanov, Andr{\'e}},
  journal={Optica},
  volume={5},
  number={9},
  pages={1150--1154},
  year={2018},
  publisher={Optica Publishing Group},
  doi = {10.1364/OPTICA.5.001150}
}

@incollection{boyd2020nonlinear,
  title={Nonlinear optics},
  author={Boyd, Robert W and Gaeta, Alexander L and Giese, Enno},
  booktitle={Springer handbook of atomic, molecular, and optical physics},
  pages={1097--1110},
  year={2008},
  publisher={Springer},
  doi = {10.1007/978-3-030-73893-8_76}
}
\end{document}